# Challenges in Grounding Language in the Real World

**Peter Lindes[1], Kaoutar Skiker[2]**

[1]Center for Integrated Cognition   [2]Conitive
peter.lindes@cic.iqmri.org  kaoutar.skiker@conitive.com

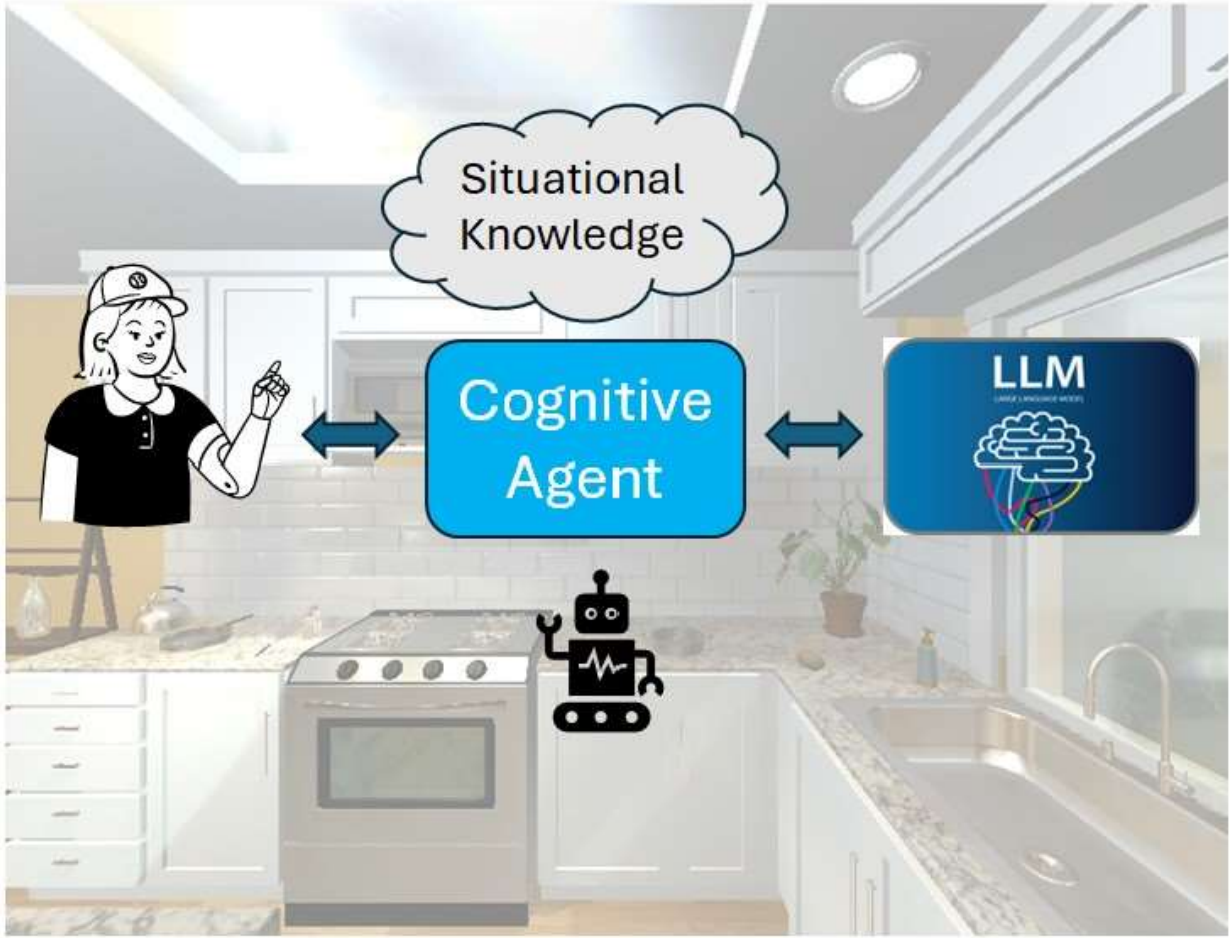

**Figure 1:** A Collaborative System that combines the memory and reasoning capabilities of a cognitive agent that controls a robot with its ability to interact with a human supervisor using language and make requests to an LLM to help it understand the human's natural language

## Abstract

A long-term goal of Artificial Intelligence is to build a language understanding system that allows a human to collaborate with a physical robot using language that is natural to the human. In this paper we highlight some of the challenges in doing this, and propose a solution that integrates the abilities of a cognitive agent capable of interactive task learning in a physical robot with the linguistic abilities of a large language model. We also point the way to an initial implementation of this approach.

# 1. Introduction

A long-term goal of Artificial Intelligence research is to develop autonomous robotic agents that can act as collaborative assistants to a human in physical, real-world environments (Nilsson, 1984). A key capability of such an agent is to understand instructions from its human partner in language that is as natural as possible to the human. This paper outlines some of the challenges in implementing such a capability, then proposes an approach that combines the strengths of an intelligent agent based on a cognitive architecture with knowledge it can extract and verify from a large language model (LLM). Finally, we describe a research program to test the viability of such a combined system.

Figure 1 is a sketch of the type of system we envision. A human supervisor communicates with a cognitive agent. That agent controls a robotic actor and maintains knowledge of the situation and the environment. It uses requests to an LLM to help it understand the human's natural language, and gather general world knowledge. Our premise is that this combined system can perform better in assisting the human than either the cognitive agent by itself or an LLM by itself.

# 2. The Language Grounding Problem

Humans are accustomed to using language to communicate with each other, including about real-world activities that a robotic physical assistant would be expected to participate in. There are several challenges in building an agent that can interact with a human partner using language as natural to the human as possible and act correctly in the world based on those interactions (Bansal et al., 2024). Consider several specific issues.

## 2.1 Identifying Objects

A human may refer to objects in the environment in many different ways. We have identified seven aspects of linguistic expressions that can be used to identify specific objects in a scene when combined with knowledge about the objects and the scene.

1. *Category* – Usually a head noun, such as *drawer* or *spoon*. These can be resolved given a list of object representations that include a category identifier, along with a mapping of the given noun onto a category identifier.

2. *Physical properties* – Adjectives such as *red*, *large*, or *square*. These can make further selection from the list of objects if the representations contain information about these properties.

3. *Deictic properties* – References based on pointing or other gestures, such as *this*, *that*, or *the one over there*. Resolving these requires knowledge of the human's pointing gestures relative to the objects in the environment.

4. *Spatial relations* – Many prepositions, such as *in*, *on*, *to the left of*, *above*, *below*, etc. refer to spatial relations. Resolving these requires knowledge of both the physical positions of objects in the scene and ways to calculate each possible linguistic reference based on these positions.

5. *Positional properties* – Many terms such as *top*, *bottom*, or *middle* also require knowledge of object positions and how to calculate whether a given object satisfies the given condition.

6. *Functional properties* – Another class of language expressions identify an object based on how it is used. Examples include *the knife drawer*, *a dessert spoon*, or *the cabinet where we keep the plates*. Resolving these requires knowledge of the situation, the kinds of tasks to be performed, what objects are needed for them, and preferences of the humans involved.

7. *Discourse properties* – Often a person in an ongoing dialog will refer to an object based on previous references in that dialog, such as *the knife* meaning the knife previously used to slice the bread. Resolving these references requires detailed knowledge of the dialog history.

## 2.2 Understanding Action Verbs in Context

Consider the following sentences using the verb *cook* and what they may mean in context.

1. *Cook.* – This could be a command to turn on the microwave to cook an object already in it.

2. *Cook the potato in the microwave.* – This describes a task that, in addition to the low-level cook operation, involves finding the potato and putting it in the microwave.

3. *Cook the potato.* – Here the task is more complicated. The agent must either know from prior knowledge or engage in dialog to find out what manner of cooking is desired using what kind of instrument.

4. *Cook an egg.* – This example emphasizes the fact that how to cook depends on what kind of object is to be cooked, and the fact that there are many ways to cook an egg. It could be hard-boiled, soft-boiled, scrambled, fried, over easy, poached, etc.

5. *Cook dinner.* – My wife will understand this sentence, as long as I say it as *Could you please cook dinner?* Our agent, however, will need a lot of background situational knowledge to know how to perform this complex task.

In general, the kinds of knowledge needed to perform a complex task labeled with the verb *cook* include at least: who is being cooked for, what kinds of objects are to be cooked, which of several possible results is desired for an item like an egg, what ingredients are required, what instrument(s) and what method(s) are needed for each type of cooking, where to find the ingredients and the instruments, and what the personal preferences of the people involved are.

## 2.3 Developing a Common Ground

A large part of understanding natural language and being able to act on it involves having common knowledge between the speaker and the hearer. For our artificial agent, this will require the ability to learn situational knowledge gradually and incrementally through its ongoing interactions with a human. This would include the sorts of knowledge mentioned above in describing referring expressions and verbs in context.

## 2.4 Understanding Free-form Language

Humans often speak in ways that are ambiguous, without situational and dialog context, are not formally grammatical, leave out assumed common knowledge, and don't state intentions clearly. We would like our agent to be able to deal with at least many of these natural variations of human speech rather than having to teach the human how to speak a restricted language understood by the agent.

# 3. Related Work

The problem of grounding language consists of mapping linguistic expressions to objects and actions in the physical world, and is fundamental to building intelligent cognitive agents (Williams et al., 2009; Harnad, 1990). Without grounding, words are only symbols, and their meanings are only defined in terms of other symbols such as in LLMs. Instead, words need to connect to and be grounded to real-world sensory and motor experiences for natural language understanding (Barsalou, 2008). Early efforts in language grounding were Winograd's (1972) SHRDLU system, that mapped natural language to objects and actions in a simulated blocks world using predefined symbolic representations. SHRDLU understands and executes commands like *put the red block on the green cube*. Although SHRDLU demonstrated how language could be grounded in a limited environment, it is unable to generalize to new environments and ambiguous language.

Subsequent research on language grounding has focused on grounding single words such as nouns, adjectives, and prepositions, as well as referring expressions that combine these elements in realistic environments. For instance, Roy and Pentland (2002) built a system where a robot learned the meanings of words like *red* and *ball* by connecting spoken words with what it saw. Tellex et al. (2011) introduced a probabilistic grounding graph that allowed robots to map spatial prepositions like *on*, *next to*, or *under* to actual locations in their environment. A robot system called INGRESS (Shridhar & Hsu, 2018) grounds expressions like *the red cup on the left* by generating descriptions for each object and comparing them to the input instruction. It handles both object attributes (like color and type) and spatial relations (like *on the left of*), and can ask clarification questions when unsure. However, it is limited to small, unconstrained environments where the set of possible referents is limited.

Grounding verbs is more difficult, as it requires understanding dynamic actions and their effects in context (see Cohen et al., 2024 for a survey on language grounding in robotics). Many approaches use parsers to transform natural language instructions into formal representations that an agent can reason over (Lindes et al. 2017). These parsers work well for well-formed commands but struggle with informal, ungrammatical, or ambiguous language, which is common in natural human communication.

Another challenge in grounding language is the establishment of a common ground—the shared knowledge and assumptions between the human and the agent about the environment and the task (Clark & Brennan, 1991). Humans rely heavily on common ground to communicate efficiently, using context-dependent expressions like *put it over there* (a deictic reference) or *do the same as before* (depends on dialog history). Without a model of shared context, an agent is likely to misinterpret such language. Lindes et al. (2017) and Thomason et al. (2015) both explored dialogue-based approaches where agents incrementally build common ground through an interactive conversation with a human.

Recently, researchers are using machine learning and deep learning techniques to address grounding in more dynamic, realistic, and complex environments. The ALFRED benchmark introduced by Shridhar et al. (2020) enabled learning from instruction-action pairs in simulated home environments. Agents learn to interpret sequences of high-level instructions and translate them into low-level actions. Similarly, Kirk et al. (2024) introduced methods for grounding ambiguous instructions using perceptual context and visual feedback. More recently, large language models (LLMs) are used to improve agents' ability to process a wide range of natural language (Ahn et al., 2022, Chen et al., 2020). For example, Ahn et al. (2022) used LLMs to map natural instructions to action sequences in real-world robot tasks in their SayCan framework. These models can bridge the gap between informal language and structured representations. However, LLMs often operate without real-time grounding in the physical environment and lack mechanisms for maintaining common ground over time. Lawley and MacClellan (2024) report on an agentic LLM system with similar goals to ours, but it lacks the abilities of a full cognitive agent.

## 4. Using an LLM to Gather Knowledge

A large language model (LLM) plays an essential part in the approach we are suggesting. It serves two primary purposes: as a source of general or common-sense knowledge, and as a tool to translate free-form language from a human into either simpler language or a formalism that the agent can understand and act on. In this section we explore briefly the idea of gathering general or common-sense knowledge from an LLM.

Previous work (Kirk et al., 2024[1], 2023) has explored using LLMs to extract knowledge needed for learning novel tasks. **Figure 2** gives a specific example of asking an LLM a high level question about how to prepare scrambled eggs.

---

[1] A video of this work is available at https://integratedcognition.ai/news/improving-knowledge-extraction-from-llms-for-task-learning-through-agent-analysis/

> **Request to ChatGPT:** (2/20/2025, 11:18am MST)
> I need a recipe for making scrambled eggs. Please give it to me as a list of a few simple steps.
>
> **Response from ChatGPT:**
> Here's a quick and easy recipe for scrambled eggs:
>
> **Ingredients:**
> - 2–3 eggs
> - Salt (to taste)
> - Pepper (to taste)
> - 1 tbsp butter (or oil)
> - Optional: Milk or cream (1–2 tbsp for creamier eggs)
>
> **Steps:**
> 1. **Crack the eggs** into a bowl and beat them with a fork or whisk until fully combined.
> 2. **Add a pinch of salt** and pepper to the eggs. Optionally, add a splash of milk or cream for a creamier texture.
> 3. **Heat a non-stick pan** over medium-low heat and add the butter.
> 4. **Pour the eggs into the pan** once the butter has melted, swirling the pan to coat the bottom.
> 5. **Cook gently**, stirring occasionally with a spatula, until the eggs are soft and slightly runny or fully cooked to your liking.
> 6. **Serve immediately** and enjoy your scrambled eggs!
>
> Let me know if you'd like to adjust the recipe to your taste!

**Figure 2:** A Recipe from ChatGPT

The result shown here would be fine for a human chef, but this language would be extremely difficult for a typical robotic language understanding system, such as that of Lindes et al. (2017), to use. Even with techniques for verifying LLM responses (Kirk et al., 2024), this level of natural language includes a very large vocabulary and its understanding depends on a great deal of additional common-sense knowledge that is not explicit in the language. Thus, we need another layer of translation to understand not only the free-form language that a human collaborator might use, but also to understand the free-form language that an LLM typically produces to answer a query about general or common-sense knowledge.

# 5. Using an LLM to Understand Free-form Language

To explore the space of the ways humans use language to describe and teach tasks, a dataset of such human instructions would make a good reference. The ALFRED dataset (Shridhar et al., 2020) has over 8,000 "demonstrations" of a robot in the AI2-Thor simulator performing one of seven types of tasks. Amazon turk "annotators" observed videos of the agent performing each task, and were asked to write both a single sentence describing the goal of the task and a detailed set of instructions based on the video to tell the agent how to perform the task. Each demonstration has such annotations made by three different human subjects.

The paragraphs below show the three annotations for a single demonstration. All three annotators saw exactly the same video. This demonstration is for a task called **Heat & Place**. This particular demonstration has a potato as the thing to be heated and a countertop as the location to place it in.

---

**Annotation # 1**
**Goal:** Put a cooked potato slice on the counter.
**Description:** Turn right, turn right, walk past the sink, turn left to face round table with tablecloth. Pick up the yellow-handled knife from the table. Cut a slice in the potato on the table. Turn left, turn left, turn right at counter, cross room, turn left at refrigerator to face counter. Put knife down on the table. Turn left, walk past sink, turn left to face round table. Pick the potato slice up from the table. Turn left, make right around corner of counter, turn left to face stove and microwave. Put potato in microwave, cook it, take it out of microwave. Turn right, cross room, turn left at counter with blue plate on it. Put potato on the counter in front of the blue plate.

---

**Annotation # 2**
**Goal:** Place a slice of cooked potato onto the counter.
**Description:** Turn right, move to the table. Pick up the knife from the table. Slice the potato on the table. Turn left, move to the counter left of the bread. Put the knife on the counter near the soap container. Turn left, move to the table. Pick up a slice of potato from the table. Turn left, move to the counter in front of the stove. Put the potato slice into the microwave, cook it, pick it back up. Turn right, move to the counter left of the bread. Put the cooked potato slice on the counter.

---

**Annotation # 3**
**Goal:** Put a piece of cooked potato on the counter.
**Description:** Turn right and cross the room, then turn left and go to face the gray table. Pick up the knife from in between the lettuce and the apple. Use the knife to slice the potato that's on the gray table. Bring the knife with you and go face the kitchen counter with the loaf of bread. Put the knife down in front of the soap dispenser on the counter. Go back over to

```
the gray table. Pick up a slice of the cut potato from the table. Bring the
potato with you and go over to the stove, then look up at the microwave. Cook
the potato slice in the microwave, then take it out again. Bring the potato
slice over to the countertop with the loaf of bread and the knife you used
to cut it. Put the potato slice down in front of the blue plate.
```

Table 1 below lists objects in the task and the different referring expressions the different annotators used to refer to them. Note that the same annotator often uses different referring expressions for the same object.

Since this data relates to the same scenes from AI2-Thor that we expect to use in our implementation, the over 25,000 ALFRED annotations can be a good source of data to test our system. These annotations are good examples of the natural "free-form" language we would like our system to understand, illustrating many different ways to refer to the same object, as well as different action verbs for the same action.

**Table 1:** Human Referring Expressions

| Object | Human1 | Human2 | Human3 |
|---|---|---|---|
| Table | Round table with tablecloth<br><br>Round table | The table | The gray table |
| Knife | Yellow handled knife | The knife | The knife from in between the lettuce and the apple<br><br>The knife<br><br>The knife used to cut the potato |
| Potato | A slice in the potato | The potato slice<br><br>The potato<br><br>A slice of potato<br><br>The cooked potato slice | The potato on the gray table<br><br>A slice of the cut potato<br><br>The potato slice |
| Counter | Counter<br><br>counter with blue plate on it | the counter near the soap container<br><br>the counter in front of the stove<br><br>the counter left of the bread | The kitchen counter with the loaf of bread<br><br>The countertop with the loaf of bread. |

To show that an LLM can translate free-form English into simpler terms, here is an example of what ChatGPT (accessed on 6/6/2025) produced for the above annotations based on this prompt, which could have been pre-programmed in the robot:

```
I am a robot trying to understand human instructions. I'm going to
give you a paragraph that a human gave me and I'd like you to
translate it into a series of short commands, one per line, with a
simple vocabulary.
```

| Annotation # 1 | | |
|---|---|---|
| Turn right | Turn left at refrigerator | Face stove and microwave |
| Turn right | Face counter | Put potato in microwave |
| Walk forward past sink | Put knife on table | Cook potato |
| Turn left | Turn left | Take potato out of microwave |
| Face round table with tablecloth | Walk past sink | Turn right |
| Pick up yellow-handled knife from table | Turn left | Walk across room |
| Cut potato on table | Face round table | Turn left at counter with blue plate |
| Turn left | Pick up potato slice from table | Put potato on counter in front of blue plate |
| Turn left | Turn left | |
| Turn right at counter | Walk forward | |
| Walk forward across room | Turn right around counter | |
| | Turn left | |

| Annotation # 2 | |
|---|---|
| Turn right | Pick up potato slice from table |
| Walk to table | Turn left |
| Pick up knife from table | Walk to counter in front of stove |
| Cut potato on table | Put potato slice in microwave |
| Turn left | Cook potato |
| Walk to counter left of bread | Take potato out of microwave |
| Put knife on counter near soap container | Turn right |
| Turn left | Walk to counter left of bread |
| Walk to table | Put cooked potato slice on counter |

| Annotation # 3 | |
|---|---|
| Turn right | Walk back to gray table |
| Walk across room | Pick up potato slice from table |
| Turn left | Walk to stove |
| Walk to face gray table | Look up at microwave |
| Pick up knife from between lettuce and apple | Put potato slice in microwave |
| | Cook potato |
| Cut potato on gray table | Take potato out of microwave |
| `Walk to kitchen counter with loaf of`<br>`bread`<br>`Face counter`<br>`Put knife in front of soap dispenser` | `Walk to counter with loaf of bread`<br>`and knife`<br>`Put potato slice in front of blue`<br>`plate` |

In Annotation #1 the human used the verb *make right*. This is unusual English, and it would be hard for the agent's parser to understand. Further interaction with the LLM improved on this:

(1)     make right around corner of counter
(2)     Make right around counter corner.
(3)     I don't understand "Make right". Can you make that simpler for me?
(4)     Walk forward
(5)     Turn right around counter

The original human sentence, (1), was originally translated as (2). A new prompt, (3), gave (4) and (5) as a result, which now the agent can understand.

These examples show that an LLM can indeed translate free-form English into a simplified form that can be easily understood by a conventional robotic comprehension system such as the one presented by Lindes et al. (2017). It also shows that when there are still things hard for the robot to understand, a further simple request can resolve that.

## 6. Our Proposed Approach

The approach we propose is to build an AI system centered on an embodied cognitive agent that interacts with both a human collaborator and an LLM, learning incrementally the situational knowledge needed to act autonomously in a complex environment (see **Figure 1**). This section describes the main components of our proposed system and how they can work together. Section 7 suggests a possible implementation.

### 6.1    Capabilities of a Cognitive Agent

Here we define a cognitive agent as an instance of a cognitive architecture with internal memories for knowledge of its situation, a way to control the perception and action of a physical robot, a means of communicating with a human in natural language, and a connection to make queries of an LLM such as ChatGPT (OpenAI, 2023).

Such an agent has the abilities to: 1) perceive the physical world, 2) perform primitive actions, 3) understand limited or formal language, 4) remember situational knowledge, 5) learn from experience, 6) do planning and reasoning, 7) have situational awareness, 8) orchestrate interactions, and 9) remember its history.

### 6.2    Capabilities of a Large Language Model

Large language models, on the other hand, have complementary abilities to: 1) handle free-form natural language, 2) ground NL to a given context, 3) translate NL to a formal representation, 4) provide general world knowledge, and 5) retain a dialog history to ground to.

## 6.3 A Combined Approach

Our primary hypothesis is that it is possible to develop a cognitive agent that can learn the large amount of situational knowledge needed to collaborate with a human on complex tasks in a complex, real-world environment from multiple interactions with a human and queries to an LLM.

The basic operation of the system begins with receiving an utterance from the human requesting an action or providing information, sending that sentence along with appropriate prompts and context information to an LLM, and receiving back a formal representation of the meaning of the sentence. Given this meaning representation, the agent will use its internal knowledge to ground the formal representation to objects in its scene, access task-related knowledge, and decide what to do next. Its next action could be adding to its knowledge base, asking for further clarification from the LLM or the human, or using its knowledge and planning capabilities to go ahead and execute the desired task in the world.

Consider one aspect of how this will work, the grounding of referring expressions in language. Section 2.1 gives seven ways of identifying objects based on language. Preliminary informal experiments that have been performed (Kirk, 2024) suggest that a modern LLM can ground language references to the agent's internal object identifiers based on object category if given a list of objects labeled with this information.

Grounding based on spatial relations and positional properties is more complicated, involving mathematical calculations on the physical position of objects in the scene. In other preliminary experiments we have found that a cognitive agent can do the math to build a graph representing neighbor relations among objects. With this graph references like *the cabinet to the left of the microwave* or the *third drawer down next to the stove* can be resolved using only symbolic calculations. The cognitive agent can easily do this resolution if given a formal representation of the reference, which the LLM should be able to provide. Alternatively, the LLM may be able to do the full resolution itself if given the neighbor graph.

Resolving functional references, such as in *Get a knife from the drawer* when there are many drawers and only one is used for storing knives, requires specific knowledge of how things are organized in this particular kitchen. Certainly, the LLM will not have that knowledge. The cognitive agent should be able to learn that knowledge either by inquiring of the human or by exploration. Once it has that knowledge, these references can be resolved easily using the agent's reasoning from the formal form of the reference.

Finally, what about references to objects previously mentioned in the dialog? Suppose the agent has used a knife to slice a potato, put the knife down somewhere to be able to move the potato slices to the pot, and now needs to get the knife again to slice an onion. Where is the knife? A cognitive agent with episodic memory should be able to remember where it left the knife. An LLM like ChatGPT (Open AI, 2023) has a context window that remembers a lot of the dialog, and it may be able to ground such reference to that. We intend to experiment with these alternatives.

Section 2.2 discussed a variety of issues related to understanding tasks based on the use of verbs in context. Human instructions such as *I'm hungry, what's for breakfast?* or even just *Cook me an egg.* require extensive situational knowledge to perform. To perform one of these actions, the agent will need knowledge of ingredients, recipes, storage locations, tools, procedures, personal preferences, and more.

Some parts of this knowledge may be gleaned by consulting the LLM, such as different ways to cook an egg. Much of it, though, can only be acquired by interaction with the human. The LLM will be an important part of this interaction to transform the free-form natural language the human uses into formal terms the agent can understand. The agent will also need its reasoning and planning abilities to verify the LLM's responses, orchestrate all these interactions, and plan the sequences of primitive actions needed to accomplish a given task.

# 7. Conclusions and Going Forward

This paper proposes an approach to tackling the problem of making a robotic agent capable of collaborating with a human in a real-world environment where much of their interaction is through natural language. The approach is based on using the proven capabilities of both cognitive agents and large language models, with each contributing its strengths to the outcome.

We have just begun to move forward with the implementation of a research prototype of the system shown in **Figure 1**. We intend to build the system in the simulated environment provided by AI2-Thor (Kolve et al., 2022), which provides a complex environment with many objects, a simulated robot that can navigate and manipulate objects in the environment, and good visuals in real time. This will be connected to an instance of the Soar cognitive architecture (Laird, 2012), which provides memories and reasoning capabilities that have been interfaced to a number of physical and simulated robots. We expect to use the dataset of natural human instructions produced using AI2-Thor in ALFRED (Shridhar et al., 2020) for learning and testing. We believe that this combination will allow us to demonstrate unprecedented abilities for a robotic agent to use natural language to collaborate with humans in real physical environments.

**Limitations**

First, we present a proposed approach to accomplishing goals of this paper, but there is no full implementation of the approach as yet. Second, the planned implementation is in simulation rather than with a physical robot. Third, we intend to work with the ALFRED dataset, but our agent will be using the symbolic data from AI2-Thor to simulate real visual perception, while their language data was produced by humans viewing a visual presentation.

## Acknowledgments

The authors wrote this paper as unfunded independent research. Many of the ideas were inspired by our work with our respective organizations.